\documentclass[twocolumn,prb,floatfix,nobibnotes]{revtex4-2}
\usepackage[utf8]{inputenc}
\usepackage{graphicx}
\usepackage[colorlinks=true,citecolor=blue]{hyperref}
\usepackage[version=4]{mhchem}

\newcommand{\td}{{\text{d}}}
\newcommand{\tn}{{\tilde{n}}}
\newcommand{\cH}{{\mathcal{H}}}
\newcommand{\bk}{{\boldsymbol{k}}}
\newcommand{\bH}{{\boldsymbol{H}}}
\newcommand{\bS}{{\boldsymbol{S}}}
\newcommand{\ha}{{\hat{a}}}
\newcommand{\hb}{{\hat{b}}}
\newcommand{\hc}{{\hat{c}}}
\newcommand{\hal}{{\hat{\alpha}}}
\newcommand{\hbe}{{\hat{\beta}}}
\newcommand{\hgam}{{\hat{\gamma}}}
\newcommand{\hze}{{\hat{\zeta}}}
\newcommand{\hmu}{{\hat{\mu}}}

\newcommand{\en}{{\hat{\boldsymbol{n}}}}
\newcommand{\ex}{{\hat{\boldsymbol{x}}}}
\newcommand{\ey}{{\hat{\boldsymbol{y}}}}
\newcommand{\ez}{{\hat{\boldsymbol{z}}}}
\newcommand{\bhe}{{\hat{\boldsymbol{e}}}}

\begin{document}
\title{Second-order correlation and squeezing of photons in cavities with ultrastrong magnon-photon interactions}
\author{Vemund Falch} 
\author{Arne Brataas}
\author{Alireza Qaiumzadeh}
\affiliation{Center for Quantum Spintronics, Department of Physics, Norwegian University of Science and Technology, NO-7491 Trondheim, Norway}
\date{March 31, 2025}

\begin{abstract}
We investigate the second-order photon correlation function in cavity-magnon systems, focusing on ferromagnetic and antiferromagnetic cavities within the ultrastrong coupling regime, and extending beyond the rotating-wave approximation. By deriving exact integral solutions for the second-order correlation function, we demonstrate that counter-rotating magnon-photon interactions induce quadrature squeezing in the cavity mode. Furthermore, we show that tuning the anisotropic magnon-cavity couplings enhances the squeezing effect by changing the level repulsion of the magnon-cavity photon hybrid mode without increasing the cavity photon occupation number. Our study reveals distinct quantum correlation behaviors in ferromagnetic and antiferromagnetic cavities: For ferromagnetic cavities, we show that squeezing increases with coupling strength asymmetry, whereas in the antiferromagnetic case, magnon modes with opposite chirality suppress quantum effects and impose a lower bound on correlation functions.
These findings provide a pathway to optimize photon blockade for quantum information technology in magnon-cavity systems in the ultrastrong coupling limit.
\end{abstract}

\maketitle

\section{\label{sec:Introduction}Introduction}
Cavity magnonics, the study of the interaction between magnons and cavity modes \cite{CavMagRev, Clerk2020, MagnonHybrids} is a promising research direction in the field of cavity quantum electrodynamics and hybrid quantum systems with potential applications in quantum information \cite{QuantumInform1}, quantum sensing \cite{QuantumSensing1, QuantumSensing2, QuantumSensing3}, and quantum transduction \cite{CavTransduction1, CavTransduction2}. Recent research has shown how magnon cavities can engineer exotic quantum condensed matter phenomena such as entangled states \cite{MagnonAtomCav, MagCav:Enta1, MagCav:Enta2,MagCav:Enta3,MagCav:Enta4} and squeezed states \cite{MagCav:Squeez1, MagCav:Squeez2} and the latter has received significant attention for their robustness against decoherence \cite{SqueezDecoherence,Hayashida2023}.

An important step in this endeavor is the realization of strong light-matter interactions by coupling cavity photons to the large number of spins in a nanomagnet \cite{StrongCoupling2, StrongCoupling1, Zhang2015, StrongMagCavCouple}. Importantly, cavities can have high quality factors \cite{MagCavCoupl5} with low damping rates reaching the strong coupling limit with high cooperativity, where the magnon-photon coupling is stronger than the damping rates \cite{Wagle_2024, MagCavCoupl4}. Recent advances have also demonstrated that the ultrastrong coupling regime, where the light-matter coupling is a \emph{sizeable} fraction of the bare frequencies, can be reached in coplanar waveguides \cite{UltraStrongCoupl}, superconducting nanostructures \cite{UltrastrongSCHy, UltraStrong2}, microwave cavities \cite{UltrastrongCoupling}, and microwave reentrant cavities \cite{BourcinUltraStrong}. Hybrid structures can also be employed to enhance the coupling strength. A recent proposal suggests utilizing topological surface states of a topological insulator to boost magnon-photon coupling \cite{Lee2023}. 
In the ultrastrong coupling regime, the counter-rotating terms, which are typically neglected in the rotating-wave approximation due to their rapid oscillation relative to the bare frequencies, become significant and cannot be ignored \cite{USCTwoLevelRes, UltraStrongCoupling2, QIN20241}. 

Antiferromagnets (AFMs) have recently received significant attention due to their THz dynamics, insensitivity to magnetic fields, and lack of parasitic stray fields \cite{Baltz:AFRev, Jungwirth:NNano2016}. Recent developments have also shown strong coupling of THz cavities to various systems \cite{USCTwoLevelRes, ThzCavity, ScalariStrongCouple}, including AFMs \cite{THzAFCavExp, Grishunin2018}. This has sparked interest in exploring the physics behind antiferromagnetic cavity magnonics, including magnon dark modes \cite{AFHam, HMagPhot, AFHam2}, magnon-magnon entanglements \cite{AFLangevin}, and enhanced superconductivity \cite{AFSuperconductor}. In addition, AFM-cavity interactions have been studied in AFMs in the GHz range \cite{JohansenPrBCAvFAF, BoventerAFCav}.  

The generation of single photons is key to enable various quantum-based technologies \cite{SinglePhotonTech1,SinglePhotonTech2}, and have been utilized for photonic quantum computation schemes \cite{SP_QC1,SP_QC2}. The current state-of-the-art single photon sources are based on semiconductor quantum dots in microcavities \cite{SP_QC1,QDCav}. A single photon state is characterized by a vanishing second-order correlation function, which measures the joint probability of finding two photons \cite{QuantumOpticsMilburn, MagnonBlockade:SCQubit}. Squeezed coherent states have long been known to exhibit sub-Poissonian statistics, which also yield a small second-order correlation function \cite{SqueezingSubPoissonian, QuantumOpticsMilburn}. Many recent proposals explore photon blockade, a phenomenon in which the presence of one photon prevents the excitation of additional photons. This effect can arise from two distinct mechanisms: conventional or unconventional blockade. In conventional blockade, strong anharmonicity causes nonuniform level spacing \cite{PhotonBlockTheoryOld, Birnbaum2005}. In the unconventional blockade, a destructive quantum interference between different excitation paths suppresses the excitation of multiple photons \cite{UncoventionalBlockadeOrig1, UnconventionalBlockadeOrig2}. The unconventional blockade in magnon-cavity systems was investigated by Ref. \cite{MagnonKerrSinglePhoton1}. Many recent works have also investigated the corresponding effect of magnon blockade, which can be used to generate single magnons, in a variety of magnon-cavity configurations with additional elements \cite{MagnonAtomCav, MagnonBlockParAmp,MagnonBlockade:SCQubit,MagnonBlockade:NonlinearCubed,MagnonBlockade:Qubit,MagnonBlockade:SCCirc}. Magnon blockade resulting from the intrinsic squeezing of magnon modes in FMs has recently been investigated in Refs. \cite{Yuan:MagnonAntibunching, WrongSqueezingApproach}.

In this paper, we perform a comprehensive investigation of the second-order photon correlation function characterizing photon blockade in ferromagnetic magnon-cavities in the ultra-strong coupling limit. We investigate how the correlations are influenced by the size and anisotropy of the rotating and counter-rotating magnon-cavity coupling strengths beyond the rotating wave approximation, and relate the results to quadrature squeezing in the cavity. The strongest blockade occurs when the amplitude of the counter-rotating magnon-cavity coupling is weaker than its rotating counterpart, which can be achieved in a circularly polarized  cavity. We note that an equivalent asymmetry in the coupling constants for magnon-magnon interactions in an AFM has also been experimentally demonstrated in Ref. \cite{Makihara2021}. The time-delayed correlations show rapid oscillations at the eigenfrequencies of the coupled systems. We then generalize our model to an antiferromagnetic magnon-cavity, where we show that the quantum effects are strongly suppressed by the presence of two magnon modes with opposite magnon chiralities.

The rest of the paper is structured as follows. First, we present the system setup and the Hamiltonian for a FM-cavity in Sec.~\ref{sec:Model}. We then solve the equations of motion for the magnon- and cavity-operators analytically in Sec.~\ref{sec:EqOfMot} by diagonalizing the system of equations in Laplace space, which is used to find an integral expression for the second-order correlation function of the cavity mode in Sec.~\ref{sec:g2Analytic}. We investigate the dependence of the equal-time correlations on the coupling constants, temperature, and detuning by computing the integral expression numerically and solving the time-delayed correlations analytically in Sec.~\ref{sec:NumRes}. We generalize our ferromagnetic model to the case of an AFM in Sec.~\ref{sec:AFM} before presenting our conclusions in Sec.~\ref{sec:Summary}.

\section{\label{sec:Model} FM-cavity Hamiltonian}
\begin{figure}
    \centering
    \includegraphics[width=0.8\linewidth]{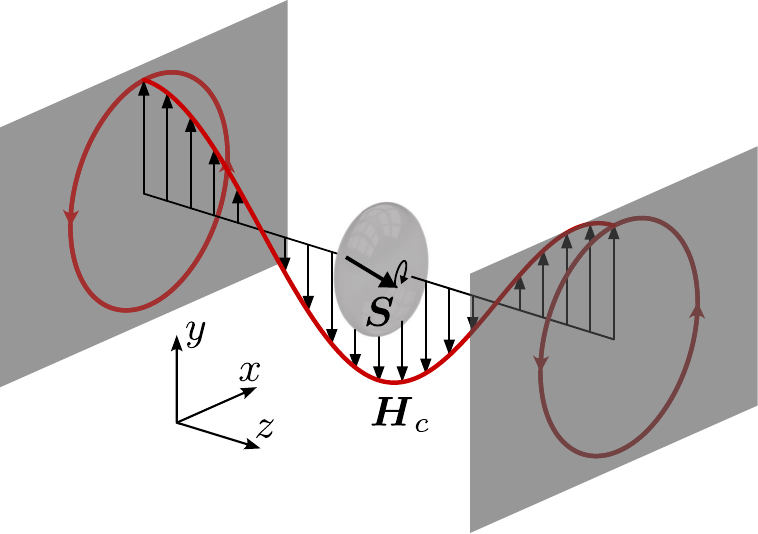}
    \caption{Schematic presentation of a simple microwave cavity consisting of two polarized conducting plates supporting a cavity mode with a circularly polarized magnetic field $\boldsymbol{H}_c(t)$, which couples though the Zeeman-interaction to the uniform magnon Kittel mode of an enclosed FM with macrospin $\boldsymbol{S}$.}
    \label{fig:CavFMFig}
\end{figure}

We consider an FM placed in a quantum electrodynamics cavity, as shown in Fig.~\ref{fig:CavFMFig}. The FM is positioned at the maximum of the magnetic field of the lowest-energy cavity mode, so that it interacts with the magnons through the Zeeman coupling. Furthermore, the magnet and cavity are independently thermalized by their own separate baths. Magnons are pumped into the FM by an externally applied coherent magnetic field. 

The total Hamiltonian of the system under study can then be separated into $\cH=\cH_{\rm c-m}+\cH_{\rm m-ba}+\cH_{\rm c-ba}+\cH_{\rm pump}$, where $\cH_{\rm c-m}$ is the Hamiltonian of the interacting cavity-magnon system, while $\cH_{\rm m-ba}$ and $\cH_{\rm c-ba}$ are the magnon-bath and cavity-bath interactions, respectively; and $\cH_{\rm pump}$ describes the coherent magnon pumping.

In our setup, the lowest-energy cavity mode predominantly interacts with the uniform Kittel magnon mode, i.e., the ferromagnetic resonance (FMR) mode. This assumption restricts the size of the FM relative to the cavity. While a larger magnet contains more spins interacting with the cavity mode, thereby enhancing the effective coupling strength, the nonuniformity of the cavity field across the magnet can also induce significant coupling with higher-order magnon modes \cite{UltrastrongCoupling}. In addition, we assume the magnet is sufficiently large to neglect higher order nonlinear terms in the Hamiltonian \cite{JohansenPrBCAvFAF}. 
Hence, the cavity quantum electrodynamics with ferromagnetic magnons can be modeled as, see Appendix \ref{secA:AppHam} for details,
\begin{align} 
\begin{split}
    \cH_{\rm c-m}=&\hbar\omega_\alpha\hal^\dagger\hal+\hbar\omega_\text{c}\hc^\dagger\hc\\
&+\hbar\big[G_r\hal^\dagger\hc+G_r^*\hal\hc^\dagger+G_n\hal\hc+G_n^*\hal^\dagger\hc^\dagger\big],\label{eq:MagCavHam}
\end{split}
\end{align}
where $\hal (\hal^\dagger)$ and $\hc (\hc^\dagger)$ are the bosonic annihilation (creation) operators for the noninteracting magnons and cavity electrodynamics modes, respectively; $\omega_\alpha (\omega_c)$ is the magnon (cavity) mode frequency, $\hbar$ is the reduced Planck constant, and $G_r (G_n)$ parametrizes the rotating (counter-rotating) magnon--cavity coupling strength, stemming from the Zeeman interaction. An asymmetry in the coupling constants $\lvert G_r\rvert\neq \lvert G_n\rvert$ can be achieved in a circularly polarized cavity, when the FM has elliptical magnon eigenmodes, for example, in the presence of an easy-plane magnetic anisotropy or by taking into account dipolar interactions. This model is thus closely related to the anisotropic quantum Rabi model \cite{AnisotropicRabiModel1,AnisotropicRabiModel3,AnisotropicRabiModel2}, where either bosonic mode is replaced with a two-level system.

The Kittel and cavity modes are each coupled to distinct thermal bosonic baths to account for the decay of their respective modes. For insulating magnets, the bath could originate from thermal phonons of the underlying lattice. Thus, we model the bath with an effective Hamiltonian that couples magnons to phonons
\cite{FMEigFrec},
\begin{equation}
\cH_\text{m-ba}=\hbar\sum_k\left[\omega_{\gamma k}\hgam_k^\dagger\hgam_k+\left(\hgam_k+\hgam_k^\dagger\right)\left(f_{\gamma k}^*\hal+f_{\gamma k}\hal^\dagger\right)\right],
\end{equation}
where $k$ denotes the bath modes with corresponding annihilation operators $\hgam_k$, $\omega_{\gamma k}$ is the bath mode frequency, and $f_{\gamma k}$ parametrizes the magnon-bath coupling.
Similarly, the cavity photon may couple to its own separate bath, e.g., the external electromagnetic environment with bosonic annihilation operators $\hze_k$,
\begin{equation}
\cH_\text{c-ba}=\hbar\sum_k\left[\omega_{\zeta k}\hze_k^\dagger\hze_k+\left(\hze_k+\hze_k^\dagger\right)\left(f_{\zeta k}^*\hc+f_{\zeta k}\hc^\dagger\right)\right],\label{eq:BathCoupleHam}
\end{equation}
where $\omega_{\zeta k}$ is the frequency of the bath modes and $f_{\zeta k}$ parametrizes the coupling. To simplify the analysis, we have assumed that both magnons and photons couple to the symmetric position operator of their respective baths. 

To drive the second-order correlations of the cavity mode into the quantum regime, a pumping term must be incorporated \cite{QuantumOpticsMilburn}. We assume magnons are pumped into the magnet by a linearly polarized magnetic field via Zeeman interactions, which lead to the following pumping Hamiltonian
\begin{equation}
\cH_\text{pump}=2\hbar\cos(\omega_p t)\big(\bar{\alpha}\hal^\dagger+\bar{\alpha}^*\hal\big),\label{eq:PumpFM}
\end{equation}
where $\omega_p$ is the frequency of the applied magnetic field, $\bar{\alpha}$ denotes the pumping parameter or coupling constant, which is proportional to the applied magnetic field amplitude and its orientation in the $x-y$ plane.

\section{\label{sec:EqOfMot}Equations of Motion}
The dynamics of the magnon and cavity modes can be derived by the Heisenberg equation of motion $\td \hat{O}/\td t=(-i\hbar)^{-1}[\cH,\hat{O}]$ \cite{QuantumOpticsMilburn,Scully_QuantumOpticsBook}. 
In this section, we focus on deriving the equation of motion for the magnon mode annihilation operator, $\hal(t)$. The derivation for the cavity annihilation operator, $\hc(t)$, follows an entirely analogous procedure and is therefore omitted for brevity. 

The equation of motion for the magnon mode dynamics reads,  
\begin{align}
\begin{split}
\frac{\td \hal(t)}{\td t}=&-i\omega_\alpha\hal(t)-iG_r\hc(t)-iG_n^*\hc^\dagger(t)\\&-2i\bar{\alpha}\cos(\omega_p t)-i\sum_kf_{\gamma k}\left[\hgam_{k}(t)+\hgam_{k}^\dagger(t)\right],
\end{split}\label{eq:EQM_init}
\end{align}
while the equation of motion for the magnon-bath dynamics is
\begin{align}\label{eq:gamma}
\begin{split}
\frac{\td \hgam_k(t)}{\td t}=-i\omega_{\gamma k}\hgam_k(t)-if_{\gamma k}^*\hal(t)-if_{\gamma k}\hal^\dagger(t).
\end{split}
\end{align}
Equation \eqref{eq:gamma} can be directly solved by integration and the result inserted back into Eq.~\eqref{eq:EQM_init} \cite{ClerkInputOutput}. To proceed, it is useful to apply the Laplace transform \cite{Ghasemian_DissipativeDyn, AdvancedEngineeringMath} and work in Laplace space with the operators $\ha_s=\mathcal{L}\{\ha(t)\}$ where $s=\Gamma+i\omega$ represents the complex frequency. The equation of motion for $\hal_s$ becomes
\begin{align}
\label{eq:FullhalEqMot_Lap}
\begin{split}
    &s\hal_s=\hal(0)-i\omega_\alpha\hal_s-iG_r\hc_s-iG_n^*\hc_s^\dagger-\frac{2i\bar{\alpha} s}{s^2+\omega_p^2}\\
    &-i\sum_kf_{\gamma k}\left(\frac{\hgam_{k0}}{s+i\omega_{\gamma k}}+\frac{\hgam_{k0}^\dagger}{s-i\omega_{\gamma k}}\right)\\
    &-\sum_k\left[\lvert f_{\gamma k}\rvert^2\hal_s+f_{\gamma k}^2\hal^\dagger_s\right]\left[\frac{1}{s+i\omega_{\gamma k}}-{\rm c.c.}\right],
    \end{split}
\end{align}
where $\hal(0)=\hal(t=0)$ and $\hgam_{k0}=\hgam_k(t=0)$ are the initial state operators of the magnon Kittel mode and the bath modes, respectively.

As demonstrated in Appendix \ref{secA:effHam}, the terms in the last line of Eq.~\eqref{eq:FullhalEqMot_Lap} can be further simplified by performing the summation over the bath modes, which requires an assumption regarding the $k$-dependent phases of the coupling constants $f_{\gamma k}$. Assuming the bath to be incoherent, we make a random phase approximation (RPA), where the phases of $f_{\gamma k}$ for different $k$ are assumed to be uncorrelated and random. Within the continuum limit, we must therefore average over all phases, leaving only contributions from terms involving $\lvert f_{\gamma k}\rvert^2$. Nevertheless, we have confirmed that for the FM-cavity system, the RPA introduces only a minor quantitative correction and does not significantly affect the results. 
In addition, we assume that the baths are characterized by an Ohmic-like spectral density. Using these assumptions, the term in the last line of Eq.~\eqref{eq:FullhalEqMot_Lap}, after summing up over all modes $k$, is reduced to an effective damping term that can be added to the magnon eigenfrequency $\omega_\alpha\to\omega_\alpha+i\eta_\alpha\omega/2$, where $\omega=\text{Im}(s)$ and $\eta_\alpha$ is the phenomenological magnon Gilbert damping parameter. By keeping the frequency dependence of the damping rate, we ensure that it is evaluated correctly at eigenfrequencies rather than the bare frequencies, which is more complicated to incorporate when using a master equation formulation \cite{MasterEqDissipation}.

Applying the above assumptions, we find the following equation of motion,
\begin{align}
    \begin{split}
         s\hal_s-\hal(0)=-\bigg(i\omega_\alpha-\frac{\eta_\alpha \omega}{2}\bigg)\hal_s-iG_r\hc_s-iG_n^*\hc_s^\dagger&\\
    -\frac{2i\bar{\alpha} s}{s^2+\omega_p^2}-i\sum_kf_{\gamma k}\left(\frac{\hgam_{k0}}{s+i\omega_{\gamma k}}+\frac{\hgam_{k0}^\dagger}{s-i\omega_{\gamma k}}\right)&.
    \end{split}\label{eq:SysEqBRWA}
\end{align}
The corresponding equation of motion for the cavity mode, $\hc_s$, can be derived in a similar way and the result is the same under applying the following substitutions: interchange the operators and indices as $\alpha\leftrightarrow c$, $\gamma\to\zeta$, replace $G_r\to G_r^*$ and set $\bar{\alpha}=0$. Note that the corresponding equation of motion for $\hal_s^\dagger$ ($\hc_s^\dagger$) is not identical to the complex conjugates of equations for $\hal_s$ ($\hc_s$) in the Laplace domain and must be separately derived from the Heisenberg equation of motion with the same procedure as above.

Finally, explicit expressions for $\hal_s,\hal_s^\dagger,\hc_s$, and $\hc_s^\dagger$ can be found by solving the resulting linear system of equations. The linear system of equations can be rewritten in a matrix form, $A[\hal_s,\hal_s^\dagger,\hc_s,\hc_s^\dagger]^{\rm{T}}={b}$, with
\begin{align}
    A&=\begin{bmatrix}
        s+i\tilde{\omega}_\alpha&0&iG_r&iG_n^*\\
        0&s-i\tilde{\omega}_\alpha&-iG_n&-iG_r^*\\
        iG_r^*&iG_n^*&s+i\tilde{\omega}_c&0\\
        -iG_n&-iG_r&0&s-i\tilde{\omega}_c
    \end{bmatrix},\label{eq:LinSys}\\
   b&=\begin{bmatrix}
       \hal(0)-\frac{2i\bar{\alpha}s}{s^2+\omega_p^2}-i\sum_kf_{\gamma k}\left(\frac{\hgam_{k0}}{s+i\omega_{\gamma k}}+\frac{\hgam_{k0}^\dagger}{s-i\omega_{\gamma k}}\right)\\
       \hal^\dagger(0)+\frac{2i\bar{\alpha}^*s}{s^2+\omega_p^2}+i\sum_kf_{\gamma k}^*\left(\frac{\hgam_{k0}^\dagger}{s-i\omega_{\gamma k}}+\frac{\hgam_{k0}}{s+i\omega_{\gamma k}}\right)\\
       \hc(0)-i\sum_kf_{\zeta k}\left(\frac{\hze_{k0}}{s+i\omega_{\zeta k}}+\frac{\hze_{k0}^\dagger}{s-i\omega_{\zeta k}}\right)\\
       \hc^\dagger(0)+i\sum_kf_{\zeta k}^*\left(\frac{\hze_{k0}^\dagger}{s-i\omega_{\zeta k}}+\frac{\hze_{k0}}{s+i\omega_{\zeta k}}\right)
   \end{bmatrix},\nonumber
\end{align}
where $\tilde{\omega}_{\alpha(c)}=\omega_{\alpha(c)}+i\eta_{\alpha(c)}\omega/2$. 
The operators $\hal_s$ and $\hc_s$ then have simple poles at the pumping frequency $s=\pm i\omega_p$, bath frequencies $s=\pm i\omega_{\gamma k},\pm i\omega_{\zeta k}$, and the complex eigenfrequencies  of the system $s=\pm i\omega_i-\Gamma_i$, $i=1,2$ obtained from $\text{det}(A)=0$.

The simple poles at the eigenfrequencies of the magnon-cavity system correspond to decaying modes that encode the transient behavior of the system. As we are only interested in the steady state regime, we work in the limit $t\Gamma_{1,2}\gg1$ where the initial state has decayed out. The exact solution to the system of equations in Eq.~\eqref{eq:LinSys} for the cavity creation operator then separates into a coherent and thermal part,
\begin{equation}
\hc(t)\vert_{t\Gamma_{1,2}\gg 1}=\Omega(t)+\hat{\mu}(t).\label{eq:cBRWA_Full}  
\end{equation}
Here $\Omega(t)$ is a time-dependent constant proportional to the strength of the pumping that encodes the coherent properties of the photon, while $\hmu(t)$ is an operator that stems from the thermal fluctuations of the baths. We can also note that including higher-order terms in the magnon-cavity Hamiltonian \eqref{eq:MagCavHam} would mix the coherent and thermal parts in Eq.~\eqref{eq:cBRWA_Full}.
It will also be useful to expand $\hmu(t)$ in terms of the bath operators,
\begin{align}
\begin{split}
\hmu(t)=\sum_{\nu=\gamma,\zeta}\sum_k\lvert f_{\nu k}\rvert\Big[e^{-i\omega_{\nu k} t} r_{\nu k}\hat{\nu}_{k0}+e^{i\omega_{\nu k} t} \tilde{r}_{\nu k}\hat{\nu}_{k0}^\dagger\Big],
    \end{split}\label{eq:hcFinishedExpr}
\end{align}
where $\nu=\gamma,\zeta$ sums over the two baths and we have split $f_{\nu k}=\lvert f_{\nu k}\rvert e^{i\phi_{\nu k}}$ to extract the amplitude factor $\lvert f_{\nu k}\rvert$, such that $ r_{\nu k}$ and $\tilde{r}_{\nu k}$ are \textit{independent} of $\lvert f_{\nu k}\rvert$ but still depending on the phase $\phi_{\nu k}$. Due to the presence of counter-rotating interactions in the Hamiltonian in Eqs.~\eqref{eq:MagCavHam} and \eqref{eq:BathCoupleHam}, the photon is squeezed and $\hmu(t)$ is a combination of creation \textit{and} annihilation operators \cite{AF_Squeezed}. Due to the ultrastrong magnon-cavity interaction assumption, $\Omega$, $r_{\nu,k}$, and $\tilde{r}_{\nu k}$ are highly nonlinear functions of the system parameters.

\section{\label{sec:g2Analytic}Second-order Correlation Function} 
The (normalized) second-order correlation function that quantifies the probability of detecting two photons at times separated by a delay $\tau$, normalized by the probability of detecting photons independently, is defined as, 
\begin{equation}\label{eq:g2}
    g^{(2)}(t,\tau)=\frac{\big\langle\hc^\dagger(t)\hc^\dagger(t+\tau)\hc(t+\tau)\hc(t)\big\rangle}{\big\langle\hc^\dagger(t+\tau)\hc(t+\tau)\big\rangle\big\langle\hc^\dagger(t)\hc(t)\big\rangle}.
\end{equation}
For equal times $\tau=0$, the quantity $g^{(2)}(t,0)$ is particularly important because it reveals the nature of the light source. We may identify three regimes: $g^{(2)}(t, 0)<1$, $g^{(2)}(t, 0)=1$, and $g^{(2)}(t,0)>1$ that indicate, respectively, sub-Poissonian statistics (single-photon sources), Poissonian statistics (coherent light sources), and super-Poissonian statistics (thermal light sources). For classical fields $g^{(2)}(t, 0)\geq1$ and so we refer to $g^{(2)}(t,0)<1$ as the quantum regime, where a photon-blockade inhibiting the population of multiple photons is characterized by $g^{(2)}(t, 0)\to0$.
In the presence of a finite time delay, $\tau \neq 0$, this quantity provides additional information about the temporal coherence and the dynamics of the light source. The cavity photons are bunched if $g^{(2)}(t, \tau) < g^{(2)}(t, 0)$, where the photons tend to appear in pairs, or are antibunched if $g^{(2)}(t, \tau)>g^{(2)}(t, 0)$.
Squeezed states can show antibunching or nonclassical boson statistics, which are directly observable through measurements of $g^{(2)}(t,\tau)$.

Inserting Eq.~\eqref{eq:cBRWA_Full} into Eq.~\eqref{eq:g2}, and using that the baths are in thermal equilibrium such that the bosonic operators $\hgam$ and $\hze$ are Gaussian since they obey Bose-Einstein statistics, we find,
\begin{align}\label{eq:g2-t}
            &g^{(2)}(t,\tau)=1+\frac{1}{g^{(1)}_{t+\tau}g^{(1)}_t}\bigg[\big\lvert\big\langle\hmu_{t+\tau}\hmu_t\big\rangle\big\rvert^2+\big\lvert\big\langle\hmu_{t+\tau}^\dagger\hmu_t\big\rangle\big\rvert^2\nonumber\\
            &+2\text{Re}\left(\Omega_t^*\Omega_{t+\tau}^*\big\langle\hmu_{t+\tau}\hmu_t\big\rangle+\Omega_t^*\Omega_{t+\tau}\big\langle\hmu_{t+\tau}^\dagger\hmu_t\big\rangle\right)\bigg],
\end{align}
where $\Omega_t=\Omega(t)$ and $\hmu_t=\hmu(t)$ for ease of readability, and $g^{(1)}_t=\langle\hc^\dagger_t\hc_t\rangle=\lvert\Omega_t\rvert^2+\langle\hmu_t^\dagger\hmu_t\rangle$ is the first order correlation function which corresponds to the occupation of the cavity mode.
At equal times, i.e., $\tau=0$, Eq.~\eqref{eq:g2-t} can be further simplified to
\begin{equation}\label{eq:g2-0}
    g^{(2)}(t, 0)=2+\frac{2\text{Re}\big[\Omega_t^{2}\langle \hat{\mu}^\dagger_t\hat{\mu}^\dagger_t\rangle\big]+\big\lvert \langle \hat{\mu}^\dagger_t\hat{\mu}^\dagger_t\rangle\rvert^2-\lvert\Omega_t\rvert^4}{\big(\lvert\Omega_t\rvert^2+\langle \hat{\mu}_t^\dagger\hat{\mu}_t\rangle\big)^2}.
\end{equation}
The expectation values in Eqs.~\eqref{eq:g2-t} and \eqref{eq:g2-0} can be compactly expressed in terms of the constants introduced in Eq.~\eqref{eq:hcFinishedExpr} by first introducing
\begin{subequations}
    \begin{align}
            \lim_{t\to\infty}\langle \hat{\mu}^\dagger_{t+\tau}\hat{\mu}_t\rangle&=\sum_{\nu=\gamma,\zeta}\big[\tilde{n}_{\nu0}(\tau)+\tilde{n}_{\nu a}(\tau)\big],\\
            \lim_{t\to\infty}\langle \hat{\mu}_{t+\tau}\hat{\mu}_t\rangle&=\sum_{\nu=\gamma,\zeta}\tilde{n}_{\nu s}(\tau)=\lim_{t\to\infty}\langle \hat{\mu}^\dagger_t\hat{\mu}_{t+\tau}^\dagger\rangle^*,
    \end{align}\label{eq:muExpNice}%
\end{subequations}
where,
\begin{subequations}\label{d-nu}
    \begin{align}
        \tilde{n}_{\nu 0}(\tau)&=\sum_ke^{i\omega_k\tau}\lvert f_{\nu k}\rvert^2\lvert r_{\nu k}\rvert^2n_{\nu k},\\
        \tilde{n}_{\nu a}(\tau)&=\sum_ke^{-i\omega_k\tau}\lvert f_{\nu k}\rvert^2\lvert \tilde{r}_{\nu k}\rvert^2\big(n_{\nu k}+1\big),\\
        \tilde{n}_{\nu s}(\tau)&=\sum_k \lvert f_{\nu k}\rvert^2r_{\nu k}\tilde{r}_{\nu k}\Big[e^{-i\omega_k\tau}\big(n_{\nu k}+1\big)+e^{i\omega_k\tau}n_{\nu k}\Big],
    \end{align}\label{eq:d_nuk}%
\end{subequations}
and $n_{\nu k}=[\exp(\hbar \omega_{\nu k}/k_\text{B}T)-1]^{-1}$ is the equilibrium Bose-Einstein distribution for mode $\nu$ at temperature $T$ with $k_\text{B}$ denoting the Boltzmann constant. Here $\tilde{n}_0=\tilde{n}_{\gamma0}+\tilde{n}_{\zeta 0}$ is the occupation of the cavity mode due to thermal fluctuations and $\tilde{n}_a=\tilde{n}_{\gamma a}+\tilde{n}_{\zeta a}$ is the occupation due to spontaneous processes that create or destroy a pair of particles, coming from the counter-rotating terms in the magnon-cavity and bath interactions. $\tilde{n}_s=\tilde{n}_{\gamma s}+\tilde{n}_{\zeta s}$ is related to the quadrature squeezing of the photon mode and must be nonzero to reach the quantum mechanical regime $g^{(2)}(t,0)<1$.

To evaluate the sums over $k$ in Eq. (\ref{d-nu}), we again use the continuum limit in the same manner as described in Appendix~\ref{secA:effHam}. The resulting integrals can then be computed numerically after inserting for the various quantities and averaging over the angles $\phi_{\nu k}$ within RPA. It can be shown that for large frequencies $\omega_k$ both $r_{\gamma k},\tilde{r}_{\gamma k}\propto\omega_k^{-2}$ drop off quickly enough that the influence of high-frequency magnon bath modes on the $\tn_{\gamma}$'s of Eqs.~\eqref{eq:d_nuk} is negligible for an Ohmic bath. However, $r_{\zeta k},\tilde{r}_{\zeta k}\propto\omega_k^{-1}$ drops off more slowly, and while $\tn_{\zeta0}$ is strongly suppressed at larger $\omega_k$'s by the thermal occupation $n_{k,\zeta}$, both $\tn_{\zeta t}$ and $\tn_{\zeta c}$ require the introduction of a cutoff frequency $\omega_\text{cut}$ for the integrals to converge. As an artifact of choosing the Ohmic baths, $\tn_{\zeta t}$ and $\tn_{\zeta c}$ are therefore strongly influenced by the high-frequency bath modes and the cutoff frequency $\omega_\text{cut}$.

\section{\label{sec:NumRes}Numerical results for the FM-cavity case}
We investigate the second-order correlation function $g^{(2)}(t,\tau)$ for the cavity mode given in Eq.~\eqref{eq:g2}. We first compute and analyze the equal-time correlations $g^{(2)}(t, 0)$ for both a time-dependent pumping field $\omega_p>0$ and a constant bias field $\omega_p=0$, before calculating the second-order correlations for a finite time delay $\tau$. 
We focus on the steady-state regime where $t\gg \Gamma_{1,2}^{-1}$, ensuring that any initial transient dynamics have decayed. For simplicity, we assume that the system has already reached the steady-state regime at $t=0$.

As indicated at the end of the Sec.~\ref{sec:g2Analytic}, the coefficients $\tn_\zeta$ in Eqs.~\eqref{eq:d_nuk}, coming from fluctuations in the cavity bath, are sensitive to both the bath cut-off frequency as well as the high-frequency behavior of the coupling constants $f_{\zeta k}$. Therefore a noisy cavity $\eta_c\gtrsim\eta_\alpha$ would give results that strongly depend on the microscopic details of the cavity bath $\zeta$, which is beyond the scope of this paper. However, as cavities with large quality factors \cite{MagCavCoupl5} whose damping rate is far below that of prototypical ferromagnets \cite{YIG_GilbertDamping1,YIG_GilberDamping2} are common, we focus on the region where $\eta_\alpha\gg\eta_c$, which is realistic for magnon-cavities \cite{UltrastrongSCHy,UltraStrongCoupl}. In the following, we set $\eta_\alpha=10^{-3}$, $\eta_c=10^{-5}$, corresponding to a high quality magnet and cavity, but we have verified that the results are qualitatively similar for larger damping rates with similar relative strengths. At temperatures $k_\text{B}T\gtrsim\hbar\omega_{1,2}$, the correlations are dominated by a significant thermal occupation and will not reach the quantum regime $g^{2}(t,0)<1$. We therefore work at low temperature, and set $k_\text{B}T=0.08\hbar\omega_\alpha$ unless specified otherwise. We furthermore choose $G_{r(n)} > 0$ for simplicity, as within the RPA any complex phase is equivalent to an in-plane rotation of the pump, i.e., the applied magnetic field.

\begin{figure}
    \centering
    \includegraphics[width=\linewidth]{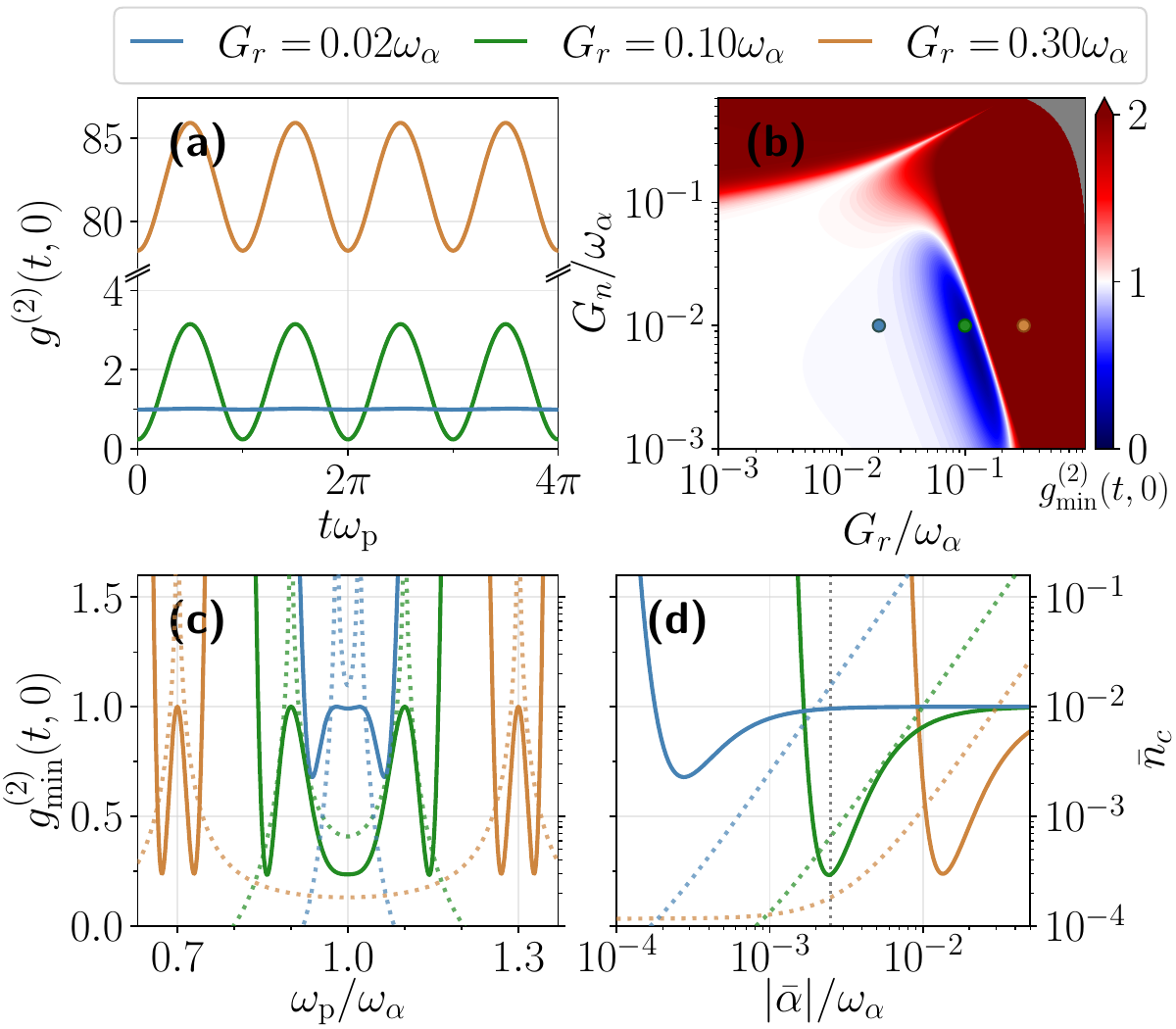}
    \caption{The equal time second-order correlation function $g^{(2)}(t,0)$ for the cavity-mode under a time-dependent drive. (a) The time dependence of $g^{(2)}(t,0)$,  which oscillates at twice the pumping frequency $\omega_p$. The minimal value of $g^{(2)}(t,0)$ over $t$ as a function of: (b) the coupling constants $G_r, G_n$, (c) the pumping frequency $\omega_{\text{p}}$ and (d) the pumping strength $\bar\alpha$. The colored dots in (b) indicate the values used in (a), (c) and (d), all for $G_n=0.01\omega_\alpha$. The dotted lines in (c) and (d) show the average cavity photon number $\bar{n}_c$. Other parameters used in the figure are $\omega_\alpha=\omega_c=\omega_p$ and $\bar{\alpha}=2.5i\times 10^{-3}\omega_\alpha$, which is indicated by the dotted line in (d).}
    \label{fig:g2FM_Eq}
\end{figure}

Figure \ref{fig:g2FM_Eq}(a) shows the equal-time correlations $g^{(2)}(t,0)$ from Eq.~\eqref{eq:g2-0} as a function of time for three different values of $G_r$, at $G_n=0.01\omega_\alpha$. The correlations oscillate at twice the pumping frequency $\omega_p$, due to the phase of the coherent part $\Omega_t$ fluctuating in time. The value of $g^{(2)}(t,0)$ is, however, very different for the three values of $G_r$. For the smallest value (blue line), the cavity exhibits near-perfect coherence. The intermediate value (green line) oscillates between the sub-Poissonian and super-Poissonian regimes, while the largest value (yellow line) remains consistently deep within the super-Poissonian regime.

We now focus on the minimum values of $g^{(2)}_\text{min}(t,0)$ over time, as we want to understand how the photons can enter the quantum regime $g^{(2)}(t,\tau)<1$. Figure~\ref{fig:g2FM_Eq}(b) shows how minima depend on the coupling constants $G_r$ and $G_n$. We can identify three different regimes corresponding to the three values in Fig. \ref{fig:g2FM_Eq}(a), which are determined by the relative size of the coherent $\lvert\Omega_t\rvert^2$ and bath-induced $\langle \hmu^\dagger\hmu\rangle$ contributions to the photon occupation number. For small values of $G_r$ and $G_n$, the coherent part is large and the photon mode is coherent $g^{(2)}(t,0)\approx1$, while for large $G_{r(n)}$ the system is in the super-Poissonian regime $g^{(2)}(t,0)>2$ as the coherent part is comparatively small. In between, there is a narrow region in the ultra-strong coupling regime where the minimum $g^{(2)}_\text{min}(t,0)$ reaches the sub-Poissonian regime, as the coherent and thermal contributions are of similar strength. The parameter-space of large $G_r,G_n\gtrsim \omega_\alpha$'s, corresponding to the gray area in Fig.~\ref{fig:g2FM_Eq}(b) where the real part of the lower eigenfrequency vanishes, is not considered in this work. This boundary is equivalent to the superradiant phase transition in the anisotropic quantum Rabi models \cite{AnisotropicRabiModel2, AnisotropicRabiModel3,AnisotropicRabiModel4}.

The location of the thin blue region in Fig.~\ref{fig:g2FM_Eq}(b) depends strongly on the pumping frequency $\omega_\text{p}$ and pumping strength $\bar{\alpha}$, as shown in Figs.~\ref{fig:g2FM_Eq}(c) and \ref{fig:g2FM_Eq}(d) respectively, where we again plot the minima in time of $g^{(2)}_\text{min}(t,0)$ (solid lines) for the same three values of $G_r$ and $G_n$ as in Fig.~\ref{fig:g2FM_Eq}(a), as well as the average cavity occupation number $\bar{n}_c=\langle\hc^\dagger(t)\hc(t)\rangle$ (dotted lines). Since both thermal averages $\langle\hmu_t^\dagger\hmu_t\rangle$ and $\langle\hmu_t\hmu_t\rangle$ are independent of $\omega_\text{p}$ and $\bar{\alpha}$, the observed behavior stems solely from the coherent part, which to lowest order in $G_r,G_n\ll\omega_\alpha$ is
\begin{equation}
    \lvert\Omega_t\rvert^2\approx\frac{\lvert \bar{\alpha}G_r\rvert^2}{\big[\Gamma_1^2+(\omega_\text{p}-\omega_1)^2\big]\big[\Gamma_2^2+(\omega_\text{p}-\omega_2)^2\big]},
\end{equation}
where $\omega_{1,2}\approx \omega_\alpha\mp G_r-G_n^2/2\omega_\alpha$ are the eigenfrequencies at $\omega_\alpha=\omega_c$. At resonance $\omega_\text{p}=\omega_i$ or for strong pumping $\alpha\gtrsim G_r$ the cavity photon is almost completely coherent with a large occupation $\bar{n}_c$ and $g^{(2)}(t,0)\approx 1$, as observed in Fig.~\ref{fig:g2FM_Eq}(c). This also causes the dispersive ``notch" at $G_n\sim0.1\omega_\alpha$ in Fig.~\ref{fig:g2FM_Eq}(b), where the upper eigenfrequency $\omega_2\approx \omega_\text{p}$ is at resonance. In the opposite limit, far from resonance or for weak pumping, $\Omega_t$ is small and so the photon is super-Poissonian $g^{(2)}(t,0)>2$ with a small thermally induced occupation $\bar{n}_c$. At some intermediate values of $\omega_\text{p}$ and $\bar{\alpha}$ where the coherent and squeezed contributions $\lvert\Omega_t\rvert^2\sim\lvert\langle\hmu_t\hmu_t\rangle\rvert\propto\lvert G_nG_r\rvert$ are similar, $g^{(2)}(t,0)$ is minimized. Thus for larger $G_r,G_n$'s stronger or more resonant pumping is required, as shown in Figs.~\ref{fig:g2FM_Eq}(c,d), and the blue sub-Poissonian region in Fig.~\ref{fig:g2FM_Eq}(b) occur in at the intermediate $G_r,G_n$'s at the transition from the coherent to super-Poissonian thermal regime. Furthermore, the occupation $\bar{n}_c$ at the pumping strength minimizing $g^{(2)}(t,0)$ will thus also increase with larger coupling strengths $G_r,G_n$, as seen in Figs.~\ref{fig:g2FM_Eq}(c,d).

For the time-dependent pumping, as discussed above, the second-order correlation function $g^{(2)}(t,0)$ is mostly governed by the behavior of the coherent part $\Omega_t$. To better elucidate the underlying physics and the origin of the sub-Poissonian cavity photons in Fig.~\ref{fig:g2FM_Eq}, we turn to the case of time-\textit{independent} ``pumping", where a static bias field is applied to the magnet. In this case $\Omega_t=\Omega$, and thus $g^{(2)}(t,0)$, are time independent in the steady-state limit. Furthermore, since the gap $\lvert\omega_\text{p}-\omega_i\rvert=\omega_i$ is already large compared to coupling strengths $G_r,G_n$, the coherent part $\Omega$ is less sensitive to $G_r$ compared to the case of time-dependent pumping discussed above.

\begin{figure}[t]
    \centering
    \includegraphics[width=\linewidth]{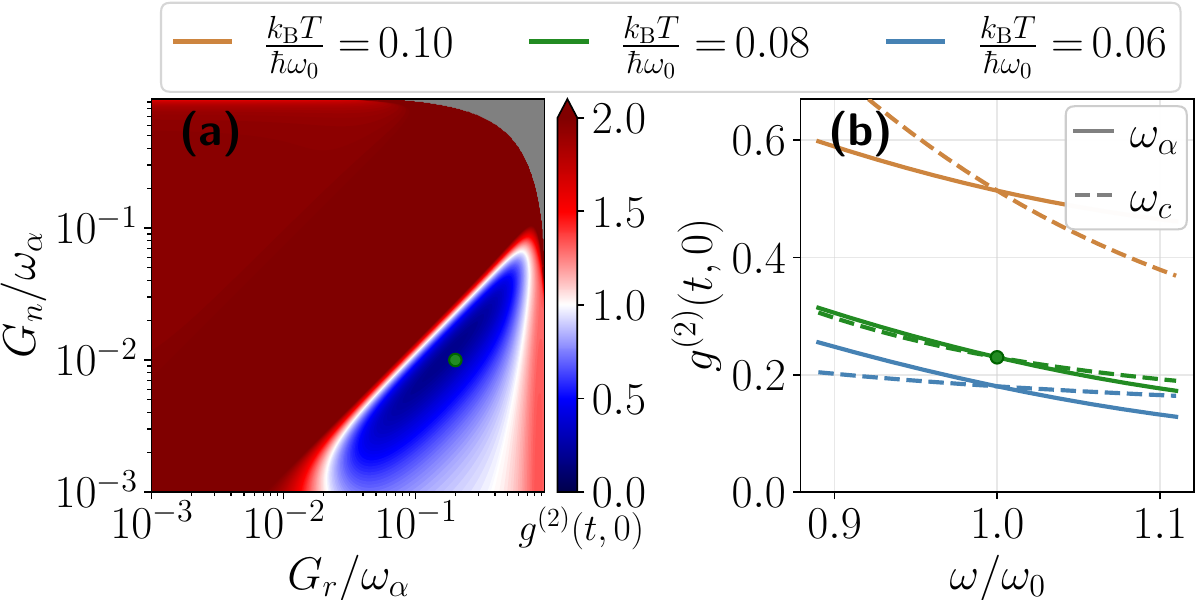}
    \caption{The equal-time second-order correlation function $g^{(2)}(t,0)$ for the cavity-mode under a constant bias field. (a) The dependence of $g^{(2)}(t,0)$ on the coupling constants $G_r$ and $G_n$ at $k_\text{B}T=0.08\hbar\omega_\alpha$ and $\bar{\alpha}=0.1i\omega_\alpha$. (b) The detuning of $\omega_\alpha$ (solid lines) or $\omega_c$ (dashed lines) from a reference-frequency $\omega_0$, equal to the bare frequencies $\omega_0=\omega_\alpha=\omega_c$ in (a), for different temperatures, where the green dot indicates the point with equal parameters in (a) and (b). The temperature and pumping strength in (b) is kept fixed at the same values relative to $\omega_0$ as in (a), and does not vary with $\omega_\alpha$ or $\omega_c$.}
    \label{fig:TimeIndep_FM}
\end{figure}

Figure \ref{fig:TimeIndep_FM}(a) shows the equal-time correlations $g^{(2)}(t,0)$ as a function of the coupling constants $G_r$ and $G_n$ for $\omega_c=\omega_\alpha$ and $\bar{\alpha}=0.1i\omega_\alpha$, corresponding to a weak applied magnetic field along the $y$ direction. A region of sub-Poissonian cavity photons $g^{(2)}(t,0)<1$ again occurs when $G_r>G_n$ in the ultra strong-coupling regime. On the diagonal $G_r=G_n$, the correlations $g^{(2)}(t,0)\geq2$ are super-Poissonian as the coherent part $\Omega=0$ vanishes. $G_r=G_n$ corresponds to where either the cavity magnetic field is normal to the bias field, i.e., linearly polarized along $\ex$, or a ferromagnet with a strong hard-axis along the bias field, i.e., $\ey$.

We now explain the occurrence of sub-Poissonian cavity modes $g^{(2)}(t,0)<1$ in Fig.~\ref{fig:g2FM_Eq} and Fig.~\ref{fig:TimeIndep_FM} in terms of quadrature squeezing \cite{QuantumOpticsMilburn}. We start by defining the quadratures $Y_1=e^{-i\phi}\hc+e^{i\phi}\hc^\dagger$ and $Y_2=-i(e^{-i\phi}\hc-e^{i\phi}\hc^\dagger)$, where $\phi$ sets the angle of the quadratures relative to the $x$- and $y$-axis, respectively. As $[Y_1,Y_2]=2i$, it follows from the uncertainty principle that the uncertainties $\Delta Y_i=\langle Y_i^2\rangle-\langle Y_i\rangle^2$ must satisfy $\Delta Y_1\Delta Y_2\geq 1$. Using Eq.~\eqref{eq:cBRWA_Full} to separate the coherent and thermal parts, we find $\Delta Y_2-\Delta Y_1=-4\text{Re}(e^{-2i\phi}\langle\hmu_t\hmu_t\rangle)$, and so the choice $\phi=\phi_0$ satisfying $e^{-2i\phi_0}\langle\hmu_t\hmu_t\rangle=-\lvert\langle\hmu_t\hmu_t\rangle\rvert$ maximizes the squeezing in favor of $Y_1$ as the minimum-uncertainty quadrature.

Pumping along the same axis as $Y_1$, i.e., $\Omega_t=\lvert\Omega\rvert e^{i\phi_0}$, such that $\langle Y_1\rangle=2\lvert\Omega\rvert$ and $\langle Y_2\rangle=0$, we can minimize the second-order correlation function
\begin{equation}
    g^{(2)}(t,0)=2+\frac{\big\lvert \langle \hat{\mu}_t\hat{\mu}_t\rangle\rvert^2-2\lvert\Omega\rvert^2\lvert\langle \hat{\mu}_t\hat{\mu}_t\rangle\rvert-\lvert\Omega\rvert^4}{\big(\lvert\Omega\rvert^2+\langle \hat{\mu}_t^\dagger\hat{\mu}_t\rangle\big)^2}.\label{eq:g2_AngleSet}
\end{equation}
Furthermore, this can only take values in the quantum regime, i.e., $g^{(2)}(t,0)<1$, when $\lvert\langle \hat{\mu}_t\hat{\mu}_t\rangle\rvert>\langle \hat{\mu}_t^\dagger\hat{\mu}_t\rangle$, which corresponds to a squeezed state along $Y_1$ as $\Delta Y_1=1+2(\langle\hmu^\dagger_t\hmu_t\rangle-\lvert\langle\hmu_t\hmu_t\rangle\rvert)<1$ and $\Delta Y_2>1$. Thus, the regions exhibiting sub-Poissonian statistics arise due to the squeezing of the system, and then pumping the photons into the quadrature with reduced uncertainty. The reduced uncertainty, in turn, causes the photon state to be more strongly peaked around the average photon number. Our choice of complex phases for magnon-cavity couplings, $G_r,G_n>0$, can correspond to a hard-axis and reduced uncertainty along the $y$ axis, we must thus orient the bias field along the same direction. We have also checked that aligning the bias field along $\ex$ gives $g^{(2)}(t,0)>1$ everywhere as expected.

We now turn to the origin quadrature squeezing in the system to understand the dependence of $g^{(2)}(t,0)$ on $G_r$ and $G_n$. Substituting Eq.~\eqref{eq:cBRWA_Full} into the magnon-cavity Hamiltonian \eqref{eq:MagCavHam}, the resulting Hamiltonian can be diagonalized by a standard Bogoliubov transformation
\begin{equation}
    \hmu=u_1\hat{\xi}_1-v_1\hat{\xi}_1^\dagger+u_2\hat{\xi}_2+v_2\hat{\xi}_2^\dagger\label{eq:BogoTransMu}
\end{equation}
in the new basis $\hat{\xi}_1,\hat{\xi}_2$, where we have dropped the temporal subscript and all the prefactors $u_i,v_i$ are chosen positive. Because of the sign difference on $v_1,v_2$, the two operators $\hat{\xi}_1,\hat{\xi}_2$ would squeeze the cavity mode along two orthogonal quadratures and their contributions cancel in the overall squeezing for $\hmu$. However, for $\omega_\alpha=\omega_c$, we find to lowest order $u_i\approx 1/\sqrt{2}$ and $v_i\approx G_n/\omega_i$. Since the interactions open a gap $\omega_2-\omega_1\approx 2G_r$ then $v_1>v_2$, and so the cancellation is only partial and $\langle\hmu\hmu\rangle\propto G_rG_n$ increases with increasing $G_r$. Furthermore, at low temperature, the thermal occupation $\langle\hat\xi_i^\dagger\xi_i\rangle=n_i\approx0$ is negligible and the occupation $\langle\hmu^\dagger\hmu\rangle\approx v_1^2+v_2^2\propto G_n^2$ is dominated by spontaneous processes which are independent of $G_r$. This leads to increased squeezing for large $G_r$'s and small $G_n$'s. Since a larger squeezing gives a smaller uncertainty, this reduces the second-order correlation function and so agrees well with the results in Figs.~\ref{fig:g2FM_Eq}(b) and \ref{fig:TimeIndep_FM}(a). However, at sufficiently large $G_r$'s, the lower eigenfrequency $\omega_1$ is so small that the thermal occupation becomes increasingly significant, while for very small $G_n$'s the thermal and pumping contributions to the occupation are larger than from the spontaneous processes, which gives a minima at finite $G_{r(n)}$.

Figure \ref{fig:TimeIndep_FM}(b) shows the dependence of the second-order correlation function on detuning the magnon frequency $\omega_\alpha$ (solid lines) or cavity frequency $\omega_c$ (dashed lines) from a reference frequency $\omega_0$, for three different temperatures. At high temperature (yellow lines) the response is dominated by a substantial thermal contribution $\langle\hmu^\dagger\hmu\rangle$, and decreasing the average frequency $(\omega_\alpha+\omega_c)/2$ is equivalent to an effective increase in the temperature, increasing $g^{(2)}(t,0)$. Furthermore, detuning the bare frequencies shifts the weights $u_i,v_i$ in Eq.~\eqref{eq:BogoTransMu} in favor of the eigenmode $\omega_i$ closest to $\omega_c$. When lowering $\omega_c$ ($\omega_\alpha$), the shift is towards the lowest (highest) frequency eigenmode with the largest (smallest) thermal occupation, causing lowering $\omega_c$ to increase $g^{(2)}(t,0)$ faster. By decreasing the temperature the thermal occupation reduces, which in turn lowers the correlation function $g^{(2)}(t,0)$. At low temperatures (blue lines) the thermal occupation is negligible and the increase for lower frequencies is instead caused by an increase in the spontaneous occupation $\tilde{n}_a$. The increase in $g^{(2)}(t,0)$ is now slower when lowering $\omega_{\alpha}$ than $\omega_c$, as the former increases the weights of the lowest frequency eigenmode which decreases the cancellation in the squeezing.

\begin{figure}[t]
    \centering
    \includegraphics[width=\linewidth]{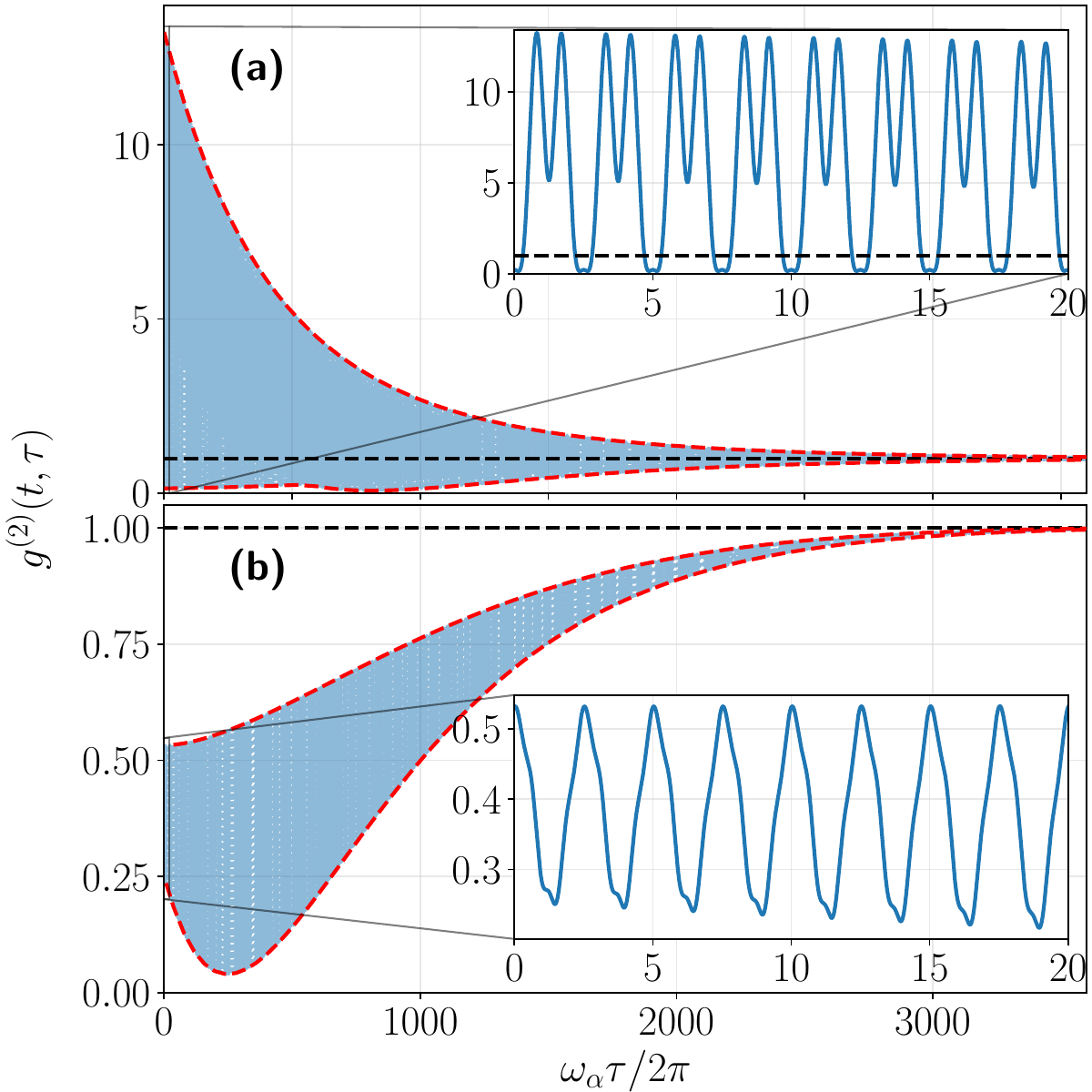}
        \caption{The time-delayed second-order correlation function $g^{(2)}(t,\tau)$ as a function of the time-delay $\tau$ for (a) $\omega_p=0$ and $\bar{\alpha}=0.1i\omega_\alpha$ and (b) $\omega_p=0.8\omega_\alpha$ and $\bar\alpha=2.5\times 10^{-5}e^{-0.1i\pi}\omega_\alpha$, where the insets show a zoomed in view for small $\tau$. The red lines showing the envelope are included as a guide to the eye. Other parameters are $G_r=0.2\omega_\alpha$, $G_n=0.01\omega_\alpha$.}
    \label{fig:g2_tau}
\end{figure}

We next turn to calculating the time-delayed correlation function for $\tau\neq 0$. Due to the rapid variation of the integrands in Eqs.~\eqref{eq:d_nuk} for large values of $\tau$, this quantity is challenging to calculate numerically. However, if the splitting between the eigenfrequencies $G_r$ is sufficiently large compared to the width of the Lorentzian $\sim\Gamma_i/2$ at each eigenfrequency as demonstrated in Appendix~\ref{secA:Loc}, the integral can be approximated analytically by treating the rest of the integrand, except the rapidly oscillating exponential, as constant over the narrow Lorentzians. In doing so, we have neglected the damping rates of the baths, which would have led to faster decay towards $g^{(2)}(t,\tau)\to 1$ as $\tau$ increases. Since we have assumed the damping rates of the baths to be larger than those of the system, see Appendix~\ref{secA:effHam}, our results are not quantitatively correct in the limit of large $\tau\gtrsim\Gamma_i^{-1}$, but should still be accurate for smaller $\tau$, which we focus on in the following.

Figure~\ref{fig:g2_tau} shows the $\tau$-dependence of the time-delayed correlation function under the above assumptions, both for a constant bias field in Fig.~\ref{fig:g2_tau}(a) and time-dependent pumping at the lower eigenfrequency $\omega_1$ in \ref{fig:g2_tau}(b). Both show a rapidly oscillating behavior as a function of $\tau$, due to the interference of the contributions from the two eigenfrequencies $\omega_1$ and $\omega_2$. However, the correlation vanishes as $g^{(2)}(t,\tau)\to1$ for larger $\tau$'s as the information is lost to the baths.

In Fig.~\ref{fig:g2_tau}(a) the correlations oscillate between $g^{(2)}(t,\tau)<1$ and $g^{(2)}(t,\tau)>1$, as the coherent part $\Omega^*_\tau=\Omega^*$ and squeezing $\langle\hmu_\tau\hmu\rangle$ in Eq.~\eqref{eq:g2-t} are out of phase. Thus, we are periodically changing between pumping along the quadratures with reduced and increased uncertainty. In Fig.~\ref{fig:g2_tau}(b) $g^{(2)}(t,\tau)<1$ at all $\tau$, as the pumping is in phase with the largest contribution since we are pumping at the lower eigenfrequency. However, the function is still oscillating rapidly as there is a significant contribution from the other eigenfrequency which is out of phase.

In both Figs.~\ref{fig:g2_tau}(a) and \ref{fig:g2_tau}(b) the correlations are also initially bunched rather than antibunched as $g^{(2)}(t,\tau)$ decreases. At equal times $\tau=0$ the cancellation in the squeezing is maximal as $v_1,v_2$ comes with different signs. However, once time-delayed they are out of phase and each picks up a phase $\exp(i\omega_i\tau)$, which reduces the degree of cancellation and thus initially lowers the $g^{(2)}(t,\tau)$.

\section{\label{sec:AFM}AFM-Cavity case} 
As we discussed in the introduction, AFMs are promising materials for spintronic-based nanotechnology since they operate at THz frequencies and have two chiral magnon modes.
To evaluate the functionality of the FM-cavity system discussed in the previous sections in comparison to their AFM counterparts, we extend our analysis to a specific model of an AFM in this section.

As shown in appendix \ref{secA:AppHam}, the general extension of FM-cavity Hamiltonian \eqref{eq:MagCavHam} to an easy-axis AFM with an applied magnetic field in the basis of the antiferromagnetic magnon modes is
\begin{align}
    \begin{split}
\cH_0=&\hbar\omega_\alpha\hal^\dagger\hal+\hbar\omega_\beta\hbe^\dagger\hbe+\hbar\omega_c\hc^\dagger\hc\\
        &+ \hbar G_1\hc\big(\hal+\hbe^\dagger\big)+\hbar G_2\hc\big(\hal^\dagger+\hbe\big)+\text{h.c.},
    \end{split}
\end{align}
where the cavity mode couples differently to magnons with opposite chirality, $\hal$ and $\hbe$, but still satisfies the sublattice symmetry $\hal\leftrightarrow\hbe^\dagger$ \cite{HMagPhot}. Pumping magnons by a linearly polarized magnetic field, the pumping Hamiltonian \eqref{eq:PumpFM} generalizes to
\begin{equation}
\cH_\text{pump}=2\hbar\cos(\omega_p t) \Big[\bar{\alpha}\big(\hal+\hbe^\dagger\big)+\text{h.c}\Big],
\end{equation} 
although the qualitative results are expected to be similar for other pumping schemes as it does not influence the squeezing of the cavity mode.

As discussed by Yuan \textit{et al.} \cite{Yuan:MasterEquation}, the two magnon modes $\hal$ and $\hbe$ may couple to different thermal baths, a common thermal bath, or both. We present here only the case for a common bath expecting it to be the most different from the FM-case, but note that the central finding should be the same for the model with two separate baths. 
If the system is invariant under exchange of the two antiferromagnetic sublattices, the Hamiltonian should be symmetric under $\hal\leftrightarrow\hbe^\dagger$, and 
we therefore consider the following effective Hamiltonian for an AFM magnon-bath system,
\begin{align}
    \begin{split}
        \cH_\text{m-ba}=&
        \hbar\sum_k\bigg\{\omega_{\gamma k}\hgam_k^\dagger\hgam_k+\left(\hgam_k+\hgam_k^\dagger\right)\\
&\times\left[f_{\gamma k}^*\left(\hal+\hbe^\dagger\right)+f_{\gamma k}\left(\hal^\dagger+\hbe\right)\right]\bigg\}.
    \end{split}\label{eq:HamAF}
\end{align}
The procedure to solve for the cavity mode operators is then analogous to the ferromagnetic case. However, in addition to introducing damping, see Appendix \ref{secA:effHam} and discussion between Eqs.~\eqref{eq:FullhalEqMot_Lap} and \eqref{eq:SysEqBRWA}, the joint magnon-bath also couples the two magnon modes, which can be modeled by adding the effective Hamiltonian
\begin{align}
    \begin{split}
        \Delta\cH_\text{AFM}=&\Big(\Delta_\alpha+i\frac{\eta_\alpha\omega}{2}\Big)\big(\hal^\dagger_s+\hbe_s\big)\big(\hal_s+\hbe^\dagger_s\big)\\
    \end{split}\label{eq:BathSplit}
\end{align}
to the equations of motion in the Laplace domain, where $\Delta_\alpha$ is a constant frequency shift induced by the bath. Since the term coupling the $\alpha$- and $\beta$-magnons only contains counter-rotating terms $\propto \hal_s^\dagger\hbe_s^\dagger+\hal_s\hbe_s$, it is expected to only give a minor correction to the result and could be removed by a suitable Bogoliubov transformation. For this reason, we set $\Delta=0$ in the following.
 
The calculation of the second-order correlation function then proceeds just as in the ferromagnetic case, where the cavity annihilation operator in the limit of large times again can be expressed as in Eqs.~\eqref{eq:cBRWA_Full} and \eqref{eq:hcFinishedExpr}, although with different constants $r_{\nu k}^{\text{nt}}$, $\tilde{r}_{\nu k}^{\text{nt}}$.

\begin{figure}[t]
    \centering
    \includegraphics[width=\linewidth]{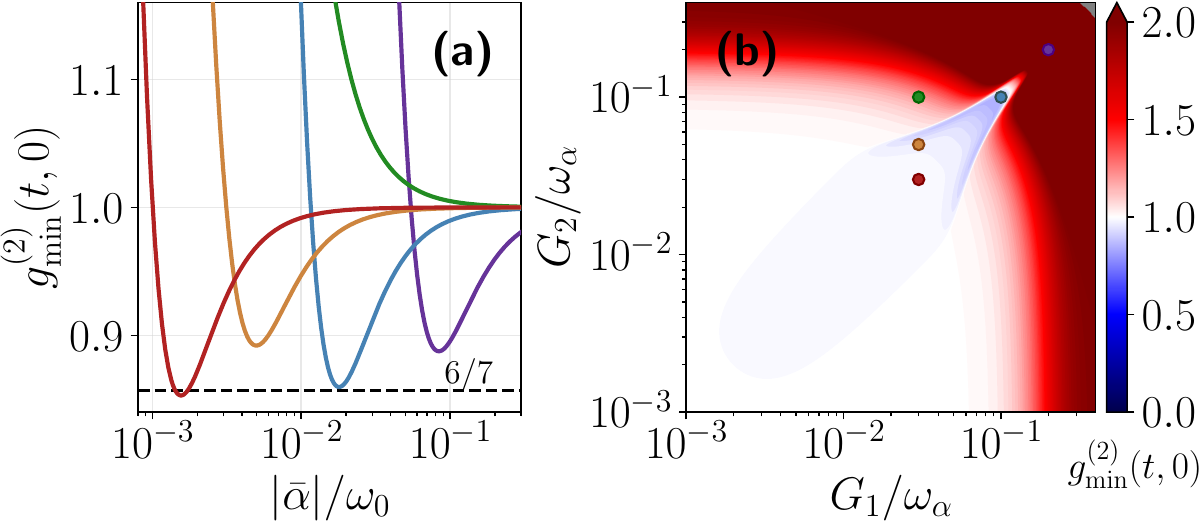}
    \caption{(a) The minimum of the equal-time second-order correlation function $g^{(2)}(t,0)$ over time $t$ for the cavity mode in an AFM-cavity, as a function of (a) the pumping-strength and (b) the coupling strengths. The values of $G_1,G_2$ in (a) correspond to the colored dots in (b), (red) $G_1=G_2=0.03\omega_\alpha$, (yellow) $G_1=0.03\omega_\alpha$, $G_2=0.05\omega_\alpha$, (green) $G_1=0.03\omega_\alpha$, $G_2=0.1\omega_\alpha$, (blue) $G_1=G_2=0.1\omega_\alpha$, (purple) $G_1=G_2=0.2\omega_\alpha$. Other parameters are $\omega_\alpha=\omega_\beta=\omega_c$, $\Delta=0$ and $\bar{\alpha}=1.5\times10^{-2}\omega_\alpha$.}
    \label{fig:g2AF}
\end{figure}

Figure~\ref{fig:g2AF}(a) and \ref{fig:g2AF}(b) show the minimum of $g^{(2)}(t,0)$ over time $t$ as a function of the pumping strength $\lvert\bar{\alpha}\rvert$ and coupling strengths $G_1$ and $G_2$, respectively. Just as for the FM-cavity in Fig.~\ref{fig:g2FM_Eq}(d), larger coupling constants require stronger pumping to achieve the minimum value of $g^{(2)}_\text{min}(t,0)$. However, the minimal values are larger than for the FM-cavity, barely dipping below $g^{(2)}(t,0)=6/7\approx0.86$, which we explain below. In Fig.~\ref{fig:g2AF} (b) we see that the minimal values now occur on the diagonal $G_1=G_2$, and that the correlations are symmetric under interchanging $G_1\leftrightarrow G_2$ due to the sublattice symmetry. 

Just as for the FM, we can diagonalize the Hamiltonian \eqref{eq:HamAF} by a suitable Bogoliubov transformation,
\begin{equation}
    \hmu=u_1\hat{\xi}_1-v_1\hat{\xi}_1^\dagger+u_2\hat{\xi}_2+v_2\hat{\xi}_2^\dagger+u_3\hat{\xi}_3+v_3\hat{\xi}_3^\dagger
\end{equation}
for some operators $\hat{\xi}_i$. Note that even though the $\hal$- and $\hbe-$magnons are coupled to the same bath, they are independent up to the rotating-wave approximation in the magnon-bath coupling, as $\langle\beta^\dagger\alpha\rangle\approx \sum f_k^2\langle\hgam_{k0}^\dagger\hgam_{k0}\rangle\to 0$ due to the RPA. For the same reason, the squeezing $\langle\hmu_t\hmu_t\rangle$ vanishes when either $G_1\to 0$ or $G_2\to 0$, which would not be the case for phase-coherent coupling, where a term $\langle\hmu_t\hmu_t\rangle\propto \sum f_k^2$ would survive.

In the limit of small coupling-constants $G_1,G_2\ll \omega_\alpha=\omega_\beta=\omega_c$ and the low-temperature limit $T\to 0$, we find $\langle\hmu\hmu\rangle\approx G_1G_2/\omega_0^2$ and $\langle\hmu^\dagger\hmu\rangle\approx(G_1^2+G_2^2)/4\omega_0^2$. Just as for the FM-cavity system, the first-order contribution to the squeezing cancels between two oppositely squeezed eigenmodes, but the opening of a gap in the eigenmodes gives a contribution to second order. The big difference is that the bath-induced occupation $\langle\hmu^\dagger\hmu\rangle\propto G_1^2+G_2^2 $ scales with both coupling-constants while the ferromagnetic result $\langle\hmu^\dagger\hmu\rangle_\text{FM}\propto G_n^2$ was given solely by the counter-rotating part. Thus in the FM-cavity the relative squeezing $\langle\hmu\hmu\rangle_\text{FM}/\langle\hmu^\dagger\hmu\rangle_\text{FM}\propto G_r/G_n$ had no upper bound to lowest order and $g^{(2)}(t,0)$ was only bound by $0$ from below. For the AFM-cavity, to lowest order in $G_1,G_2$ the relative squeezing is bounded by $\langle\hmu\hmu\rangle_\text{AFM}/\langle\hmu^\dagger\hmu\rangle_\text{AFM}\approx4G_1G_2/(G_1^2+G_2^2)\leq 2$, which corresponds to a $g^{(2)}(t,0)\geq 6/7$ in Eq.~\eqref{eq:g2-0}, with the minimum occurring on the diagonal $G_1=G_2$. This explains the observed behavior in 
Fig.~\ref{fig:g2AF} (a) and (b), even though the bound is not exact due to higher-order contributions in $G_1,G_2$.

\section{\label{sec:Summary}summary and concluding remarks}
We have studied photon second-order correlations and squeezing through cavity quantum
electrodynamics for both FM and AFM cavity systems. Going beyond the rotating-wave approximation in the ultra-strong coupling regime, we have investigated how the size and anisotropy of the rotating and counter-rotating magnon-cavity coupling terms influence the second-order correlation function. The presence of both leads to quadrature squeezing of the cavity mode, reducing the uncertainty along one of the quadratures. By pumping the system along the quadrature with reduced uncertainty, the cavity-mode can exhibit sub-Poissonian statistics characteristic of a photon blockade, although the system is still bunched rather than antibunched. Moreover we show how ferromagnetic and antiferromagnetic cavities are qualitatively different as the sublattice symmetry in an AFM puts a lower bound on the second-order correlation function.

The second-order correlation function could be measured directly by a Hanbury-Brown-Twiss experiment \cite{HBS} as all parameter choices should be experimentally feasible, which has already been done for photon blockades in other hybrid cavity systems in the weak driving regime \cite{ObservationBlock2,UPB3,UPB4}. However, due to the increasing availability of the ultra-strong coupling regime in magnon-cavities, it would be also interesting to study the interplay between anisotropic couplings and Kerr nonlinearities \cite{USCTwoLevelRes, MagnonKerrSinglePhoton1} or parametric amplification \cite{MagnonBlockParAmp} conventionally used to generate blockades, and potentially optimize the blockade efficiency. Our result are also instructive for understanding the fundamental role of intrinsic squeezing in magnon-cavity systems, with applications for decoherence robust quantum technology \cite{SqueezDecoherence,Hayashida2023}.

\section*{Acknowledgements}
We thank Jeroen Danon for useful discussions and a critical read-through of the manuscript.
The Research Council of Norway has supported this work through its Centers of Excellence funding scheme, project number 262633, "QuSpin."

\appendix
\section{Hamiltonian of free magnons and cavity subsystems}\label{secA:AppHam}
\emph{Ferromagnetic Hamiltonian --} 
The Hamiltonian for an exchange-coupled ferromagnet with an easy-axis along the $z$ direction, a hard axis along the $y$ direction, and with an applied spatially uniform magnetic field $\bH(t)=H_0\ez+H_x(t)\ex+H_y(t)\ey$ (in units of Tesla), can be expressed as \cite{HahnKopietzFMHam, JohansenPrBCAvFAF,Yuan:MagnonAntibunching}
\begin{align}
    \begin{split}
            &\cH_\text{FM}=\sum_{i,j}\sum_{\alpha\beta}\big(J_{ij}\delta_{\alpha\beta}+D_{ij}^{\alpha\beta}\big)S_i^\alpha S_j^\beta\\
            &\hspace{0.4cm}-K_z\sum_i (S_i^z)^2+K_y\sum_i (S_i^y)^2-\hbar\gamma\sum_i\bH\cdot\bS_i
    \end{split}\label{eqA:FMHamRealSpace}
\end{align}
where $i,j$ labels the sites of the lattice with spins $S^{\alpha/\beta}_{i/j}$ whose Cartesian components are labeled by $\alpha,\beta$, $J_{ij}$, and $D_{i,j}^{\alpha}$ are the matrix elements of the exchange and dipolar interaction respectively, $K_z>0$ and $K_y>0$ are the easy- and hard-axis anisotropy coefficients, and $\gamma$ is the gyromagnetic ratio.

Through a Holstein-Primakoff transformation, the Hamiltonian \eqref{eqA:FMHamRealSpace} can be recast in terms of the magnon creation and annihilation operators $\ha_i,\ha_i^\dagger$ \cite{Yuan:MagnonAntibunching, BoventerAFCav}. After performing a Fourier transformation into a momentum space $\bk$, we choose to focus exclusively on the uniform mode with momentum $\bk=0$ as outlined in the main text. In this basis, the Hamiltonian \eqref{eqA:FMHamRealSpace} takes the form   
\begin{align}
    \begin{split}
        \cH_\text{FM}=&A_\text{FM}\ha^\dagger\ha+\frac{B_\text{FM}}{2}\big(\ha\ha+\ha^\dagger\ha^\dagger\big)\\
        &+\hbar\gamma\sqrt{\frac{SN}{2}}\Big\{\big[H_x(t)-iH_y(t)\big]\ha+\text{h.c}\Big\},\label{eqA:FMHamUniform}
    \end{split}
\end{align}
where $N$ is the number of spins and
\begin{subequations}
    \begin{align}
        \begin{split}
            A_\text{FM}=&\hbar\gamma H_0+2K_zS+SK_y,\\+&S\bigg(D_0^{zz}-\frac{D^{xx}_0+D_0^{yy}}{2}\bigg),
        \end{split}\\
        B_\text{FM}=&-SK_y-S\frac{D^{xx}_0-D_0^{yy}}{2},
    \end{align}%
\end{subequations}
where $D_0^{ii}$ are the diagonal elements of the Fourier-transformed dipolar tensor \cite{HahnKopietzFMHam}. The Hamiltonian \eqref{eqA:FMHamUniform} can be diagonalized by a Bogoliubov transformation $\hal=u_\text{FM}\ha+v_\text{FM}\ha^\dagger$, where \cite{FMEigFrec}
\begin{equation}
    u_\text{FM}\big(\lvert v_\text{FM}\rvert\big)=\sqrt{\frac{A_\text{FM}+(-) \hbar\omega_\text{FM}}{2\hbar\omega_\text{FM}}},
\end{equation}
$\text{sgn}(v_\text{FM})=\text{sgn}(B_\text{FM})$, such that in the $\hal$-basis $\cH_\text{FM}=\hbar\omega_\text{FM}\hal^\dagger\hal+\bar{\alpha}(t)\hal^\dagger+\bar{\alpha}^*(t)\hal$ with $\hbar\omega_\text{FM}=\sqrt{A^2_\text{FM}-B^2_\text{FM}}$ and
\begin{align}
    \begin{split}
        \bar{\alpha}^*(t)=\sqrt{\frac{SN}{2}}\Big\{u_\text{FM}\big[H_x(t)-iH_y(t)\big]&\\-v_\text{FM}\big[H_x^*(t)-iH_y^*(t)\big]&\Big\}.
    \end{split}
\end{align}

\emph{Antiferromagnetic Hamiltonian --} A simple AFM with nearest neighbor exchange-interaction, an externally applied magnetic field, and an easy-axis magnetic anisotropy can be described by the Hamiltonian
\begin{equation}
    \cH_\text{AF}=J\sum_{\langle i,j\rangle}\bS_i\cdot\bS_j-\hbar\gamma\sum_i\bH\cdot\bS_i-K_z\sum_{i}(S_i^z)^2,\label{eqA:AFHamRealSpace}
\end{equation}
where $K_z>0$ is the easy axis anisotropy coefficient. For simplicity we do not include pumping in this derivation, as it is straightforward to generalize from the FM-case.
Again performing the Holstein-Primakoff transformation and Fourier transformation, and concentrating on the uniform $\bk=0$ mode the Hamiltonian \eqref{eqA:AFHamRealSpace} becomes \cite{HMagPhot, AFLangevin}
\begin{equation}
    \cH_\text{AF}=A_\text{AF}^a\ha^\dagger\ha+A_\text{AF}^b\hb^\dagger\hb+B_\text{AF}\big(\ha\hb+\ha^\dagger\hb^\dagger\big),\label{eqA:AFHamOp}
\end{equation}
where
\begin{subequations}
    \begin{align}
    A_\text{AF}^{a(b)}&=SJz+2SK_z\pm\hbar \gamma H_0\label{eqA:AFHamA},\\
    B_\text{AF}& = SJz,
    \end{align}
\end{subequations}
and $z$ is the number of nearest neighbors.
The Hamiltonian \eqref{eqA:AFHamOp} can be diagonalized by the Bogoliubov-transformation
\begin{equation}
\hal=u_\text{AF}\ha+v_\text{AF}\hb^\dagger,\hspace{1cm} \hbe = u_\text{AF}\hb+v_\text{AF}\ha^\dagger\label{eqA:AF_Bogo}
\end{equation}
where
\begin{subequations}
\begin{align}
    u_\text{AF}\big(v_\text{AF}\big)&=\sqrt{\frac{\bar{A}_\text{AF}\pm\hbar\bar{\omega}}{2\hbar\bar{\omega}}},
\end{align}
\end{subequations}
$\bar{A}_\text{AF}=(A_\text{AF}^a+A_\text{AF}^b)/2$ and $\hbar\bar{\omega}=\sqrt{\bar{A}_\text{AF}^2-B_\text{AF}^2}$, such that in the $\hal,\hbe$-basis $\cH_{\text{AFM}}=\hbar\omega_\alpha\hal^\dagger\hal+\hbar\omega_\beta\hbe^\dagger\hbe$ where
$\hbar\omega_{\alpha(\beta)}=\hbar\bar{\omega}\pm \hbar\gamma H_0$.

\emph{Cavity Hamiltonian --} We assume that the magnon modes couple strongly to a single optical cavity mode $\hc$ with frequency $\omega_c$. To derive the relevant coupling constants, we consider a simplified geometry where the microwave cavity is enclosed by two perfect conductors, so that the cavity is quantized along the $\en$-direction \cite{JohansenPrBCAvFAF}. The magnetic field of the cavity mode can then be described by \cite{JohansenPrBCAvFAF,QuantumOpticsMilburn}
\begin{equation}
    \bH_c=i\sqrt{\frac{\mu_0\hbar\omega_c}{V}}\big[\hc(\en\times\bhe_c)-\hc^\dagger(\en\times\bhe_c^*)\big]\cos(\bk_n\cdot\boldsymbol{r}),
\end{equation}
where $\bhe_c$ is the polarization unit vector of the cavity mode.

For simplicity we orient the coordinate system so that the magnetic quantization-axis is oriented along the $\ez$-direction. For a circularly polarized cavity-mode $\bhe_c=(\ex+i\ey)/\sqrt{2}$ we find that the Zeeman-coupling  for the AFM becomes \cite{HMagPhot, AFLangevin}
\begin{align}
    \begin{split}
        \cH_\text{int}&=-\hbar\gamma\sum_i\bH_c\cdot\bS_i\\
        &=\hbar g_0\Big[\hc\big(\ha+\hb^\dagger\big)+\text{h.c.}\Big],\label{eqA:Zeeman}
    \end{split}
\end{align}
where \cite{JohansenPrBCAvFAF}
\begin{equation}
g_0=-\gamma\sqrt{\frac{\hbar\omega_c \mu_0SN_\text{AF}}{V}}\zeta,
\end{equation}
$N_\text{AF}$ is the number of unit cells in the AF, and $\zeta=N_\text{AF}^{-1}\sum_{i\in\text{AF}}\cos(k_zz_i)$ is an overlap factor between the magnet and the cavity-field with $z_i$ the $z$-position of the unit-cells in the AF.
The result for the opposite polarity is found by replacing $\ha\leftrightarrow\hb$. The results for the ferromagnet can be found by removing all $\hb$'s from the antiferromagnetic versions.

Under the antiferromagnetic Bogoliubov transformation \eqref{eqA:AF_Bogo}, the effective coupling constant $g_0\to(u_\text{AF}-v_\text{AF})g_0<g_0$ is reduced from its bare value, vanishing at large squeezing. Furthermore, the sublattice interchange symmetry $\hal\leftrightarrow\hbe^\dagger$ is preserved. For a more complex AFM, e.g., including a hard-axis magnetic anisotropy, the Bogoliubov transformation \eqref{eqA:AF_Bogo} would also include terms coupling $\alpha$ to $\alpha^\dagger,\beta$ \cite{JohansenPrBCAvFAF,FMHam3,Shiranzaei_2022}, giving rise to new terms $\propto\hc(\hal^\dagger+\beta)$ and their complex conjugates.

Inserting the Bogoliubov transformation for the ferromagnet, we instead find
\begin{equation}
\hc\ha+\hc^\dagger\ha^\dagger\to\hc(u_\text{FM}\hal-v_\text{FM}\hal^\dagger)+\text{h.c.},
\end{equation}
which leads to an asymmetry in the strength of the rotating and counter-rotating coupling terms when $u_\text{FM}\neq v_\text{FM}$. Note that at $\bk=0$ the Bogoliubov-transformation both $u_\text{FM},v_\text{FM}$ are real, and as such the phase-difference between rotating and counter-rotating term must be $i\pi n$.

\emph{Static Pumping}
In principle, when applying a static magnetic field orthogonal to the chosen quantization direction for the magnets, a new, slightly tilted, quantization axis should be chosen instead. This would lead to a finite in-plane static magnetization, which through the Zeeman coupling in Eq.~\eqref{eqA:Zeeman} would give rise to a static pumping term for the cavity. Such a static pumping term for the cavity could equivalently be achieved by rotating the magnet slightly.  However, if the biasing field $\bar{\alpha}$ is sufficiently small, the above approximation should be reasonable and give results that are applicable to both the case of static as well as a time-dependent biasing field.

\section{Finding the effective bath-induced Hamiltonian}\label{secA:effHam}
Here we perform the sum over the ferromagnetic bath modes $k$ with frequencies $\omega_{\gamma k}$ in the last line of Eq.~\ref{eq:FullhalEqMot_Lap} to show how to find the effective damping rate for the magnon mode in Eq.~\eqref{eq:SysEqBRWA}.

\begin{figure}[t!]
    \centering
    \includegraphics[width=\linewidth]{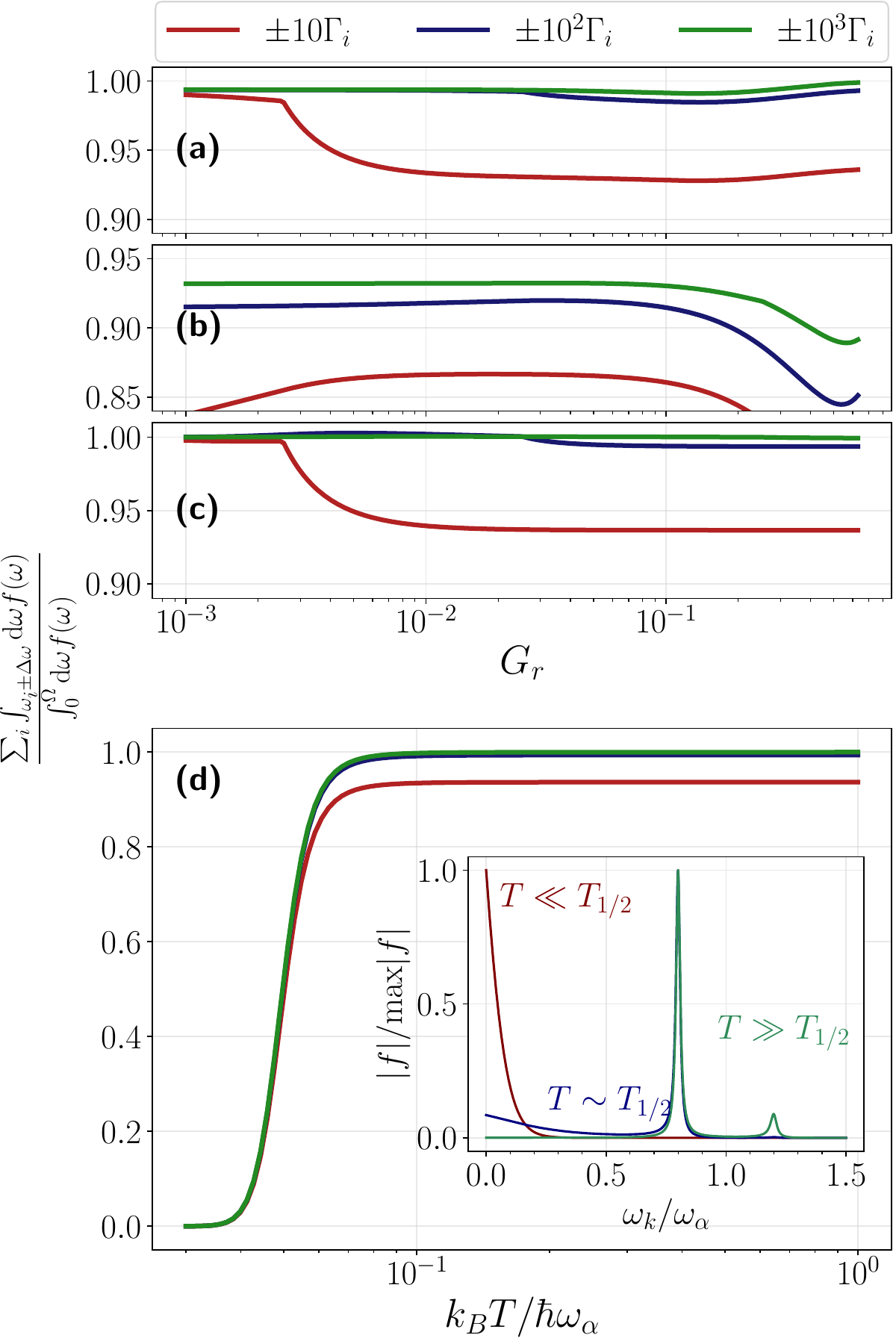}
    \caption{The localization of the integrals (a) $\tilde{n}_0$, (b) $\tilde{n}_a$ and (c) $\tilde{n}_s$ from Eq.~\eqref{eq:d_nuk} for $G_n=0.01\omega_0$ and $k_\text{B}T=0.08\hbar\omega_0$ as a function of $G_r$, and (d) $d_0$ as a function of $T$ for $G_r=0.2\omega_0$ and $G_n=0.01\omega_0$, with the inset showing the integrand for different temperature, where the damping rates have been exaggerated for clarity. The localization is computed as the ratio of the integral in a set region around the eigenfrequencies to the full integral.} 
    \label{fig:localization}
\end{figure}

The system of equations has poles at $s=i\omega+\Gamma$, where the real frequency $\omega$ attains positive or negative values for the rotating and counter-rotating parts, respectively. The damping rate of the bath modes $\Gamma<0$ must be negative to reflect loss of information to the bath. Inserting for the pole at $s=i\omega+\Gamma$, we find
\begin{align}
\begin{split}
    &\sum_k \lvert f_{\gamma k}\rvert^2\left(\frac{1}{s+i\omega_{\gamma k}}-\frac{1}{s-i\omega_{\gamma k}}\right)\label{eqA:DefDamp}\\
    &=-\frac{i}{\pi}\int \td \omega_k\,J_{\gamma}(\omega_k)\frac{\omega_k}{\omega_k^2-\big[\omega-i(\Gamma_\gamma+\Gamma)\big]^2},\\
\end{split}
\end{align}
where we have introduced the spectral density $J_\gamma(\omega)=2\pi\sum_k\lvert f_{\gamma k}\rvert^2\delta(\omega-\omega_k)$ and we have introduced a damping rate $\Gamma_\gamma$ for the bath, as information is also lost in the bath. Next, assuming an Ohmic bath $J_\gamma(\omega)=\eta_\alpha\omega/(1+\omega^2/\Omega_\gamma^2)$ with cutoff-frequency $\Omega_\gamma$ \cite{TakeiKeldysh} and assuming that $\Omega_\gamma\gg\omega$ and $\Gamma\ll\omega$, the integral in Eq.~\eqref{eqA:DefDamp} can be computed to find that we must modify the equation of motion \eqref{eq:FullhalEqMot_Lap} by changing
\begin{equation}
    i\omega_\alpha\to i\Big[\omega_\alpha -\eta_\alpha\frac{\Omega-2(\Gamma_\gamma+\Gamma)}{4}\Big]-\frac{\eta_\alpha}{2}\omega\label{eqA:DampingAndFreqShift}
\end{equation}
under the Markov approximation $\Gamma_\gamma\gg\lvert\Gamma\rvert$. It should be noted that the real dissipative part switches sign when $\Gamma_\gamma<-\Gamma$. However, it can be shown that in this case the system of equations of motion has no poles as the damping rate in the equations of motion is opposite that of $s$.

The constant frequency-shift $\Delta_\alpha=-\eta_\alpha(\Omega-2\Gamma_\gamma)/4$ due to the cutoff $\Omega$ and damping rate $\Gamma_\gamma$ can be subsumed into the definition of a renormalized magnon-frequency, while we will neglect the small frequency shift $\eta_\alpha\Gamma/2$, with $\Gamma=\text{Re}(s)$, other than noting that it ensures that the function is holomorphic around the poles. The damping rate can then be identified as $\eta_\alpha\omega/2$, and it can be verified that for the pole the frequency $s=-i\omega_\alpha$ the damping-rate is negative, as required. 

\section{\label{secA:Loc}Localization}

Figures~\ref{fig:localization}(a)-(c) shows to which degree the integrals in Eqs.~\eqref{eq:d_nuk} are localized close to the eigenfrequencies $\omega_i$ in the continuum limit. As seen, they are mostly localized at the eigenfrequencies of the magnon-cavity system. $\tilde{n}_a$ in Figure~\ref{fig:localization}(b) is the least localized, as it originates counter-rotating terms in the magnon-bath Hamiltonian \eqref{eq:BathCoupleHam} that do not couple resonantly to the magnons. Figure~\ref{fig:localization}(d) shows that for low temperatures, the thermal occupation is dominated by low-frequency modes, due to the occupation of the resonant modes being too small.

\bibliography{references}

\end{document}